\documentclass{amsproc}
\usepackage{amssymb,amscd}
\usepackage{graphics}
\usepackage{epsfig}

\theoremstyle{definition}

\theoremstyle{remark}

\numberwithin{equation}{section}

\catcode`\@=11
\def\@citex[#1]#2{%
\if@filesw \immediate \write \@auxout {\string \citation {#2}}\fi
\@tempcntb\m@ne \let\@h@ld\relax \def\@citea{}%
\@cite{%
  \@for \@citeb:=#2\do {%
    \@ifundefined {b@\@citeb}%
      {\@h@ld\@citea\@tempcntb\m@ne{\bf ?}%
      \@warning {Citation `\@citeb ' on page \thepage \space undefined}}%
      {\@tempcnta\@tempcntb \advance\@tempcnta\@ne%
      \@tempcntb\number\csname b@\@citeb \endcsname \relax%
      \ifnum\@tempcnta=\@tempcntb %Number follows previous--hold on to it
        \ifx\@h@ld\relax%
          \edef \@h@ld{\@citea\csname b@\@citeb\endcsname}%
        \else%
          \edef\@h@ld{\ifmmode{-}\else--\fi\csname b@\@citeb\endcsname}%
        \fi%
      \else%   %  non-successor--dump what's held and do this one
        \@h@ld\@citea\csname b@\@citeb \endcsname%
        \let\@h@ld\relax%
      \fi}%
    \def\@citea{,\penalty\@highpenalty\,}%
  }\@h@ld
}{#1}}

\def\@citeb#1#2{{[#1]\if@tempswa , #2\fi}}
\def\@citeu#1#2{{$^{#1}$\if@tempswa , #2\fi }}
\def\@citep#1#2{{#1\if@tempswa , #2\fi}}

\newcommand{\beqa}{\begin{eqnarray}}
\newcommand{\eeqa}{\end{eqnarray}}

\newcommand{\Z}{{\mathbb Z}}

\newcommand{\R}{{\mathbb R}}
\newcommand{\C}{{\mathbb C}}
\newcommand{\PP}{{\mathbb P}}
\newcommand{\e}{\,{\rm e}}
\newcommand{\CP}{{\C\PP}}

\newcommand{\bQ}{\overline{Q}}
\newcommand{\Tr}{{\rm Tr}}
\newcommand{\bartial}{\overline{\partial}}

\begin{document}

\title{Mirror Symmetry And Quantum Geometry}

\author{Kentaro Hori}
%\address{Department of Mathematics, Louisiana State University, Baton
%Rouge, Louisiana 70803}
\curraddr{Institute for Advanced Study, 
Princeton, NJ 08540, U.S.A.;
University of Toronto,
Toronto, Ontario M5S 1A7, Canada.}
\email{hori@ias.edu}

%    General info
%\subjclass{}

%\keywords{}

\begin{abstract}

Recently, mirror symmetry is derived as
T-duality applied to gauge systems that flow to non-linear sigma models.
We present some of its applications to study quantum geometry involving
D-branes.
In particular, we show that one can employ
D-branes wrapped on torus fibers
to reproduce the mirror duality itself,
realizing the program of Strominger-Yau-Zaslow in a slightly different
context.
Floer theory of intersecting Lagrangians plays an essential role.

%\\[0.3cm]
%{\bf 2000 Mathematics Subject Classification:}
%81T30,81T60,14J32,53D40,53D45\\
%{\bf Keywords and Phrases:}
%Mirror symmetry, D-branes, Floer homology.
\end{abstract}

\maketitle

\vspace{-0.5cm}
\hspace{3cm}
{\it To the memory of Sung-Kil Yang}
\vspace{0.5cm}

%%%%%%%%%%%%%%%%%%%%%%%%%%%%%%%%%%%%%%%%%%%%%%%%%%%%%%%%%%%%%%%%%%%%%%%%
\section{Introduction}

Mirror symmetry has played
important roles in exploring quantum modification of geometry
in string theory.
Things started with the discovery of mirror pairs of Calabi-Yau manifolds
\cite{GP} with a subsequent application \cite{Candelas} to
superstring compactifications.
It had an immediate impact on enumerative geometry and
motivated various mathematical investigations
including the formulation of Gromov-Witten invariants.
Another breakthrough was made through the
recognition of D-branes as indispensable elements in string theory
\cite{polchinski},
which was preceded by Konstevich's homological mirror symmetry
\cite{Kon}.
Studies and applications of mirror symmetry involving D-branes
have enriched our understanding of quantum geometry.

In particular,
Strominger-Yau-Zaslow (SYZ) proposed, using the transformation
of D-branes under T-duality, that mirror symmetry of Calabi-Yau manifolds
is nothing but dualization of special Lagrangian torus fibrations
\cite{SYZ}.
This provides a very geometric picture of mirror symmetry
that inspired many physicists and mathematicians.

Recently, another progress is made
via an exact analysis of
quantum field theory on the worldsheet \cite{HV}. 
Mirror symmetry is derived as T-duality applied to gauge systems
\cite{phases}
that flow to non-linear sigma models.
This, however, turns sigma models into Landau-Ginzburg (LG) models,
where the LG potential for the dual fields
is generated by the vortex-anti-vortex gas
of the high energy gauge system.

What is the relation of this
and SYZ, both of which use T-duality applied to torus fibers?
Since SYZ employ D-branes wrapped on the torus fibers, it is a natural idea
to do the same.
In this talk, we present some applications of \cite{HV}
to study the properties of D-branes.
In particular,
we show that the study of D-branes wrapped on torus fibers
indeed reproduces the LG mirror of \cite{HV}.
Study of Floer homology for intersecting Lagrangians
\cite{FOOO} plays an important role.
We will also present other aspects of mirror symmetry
involving D-branes.

%%%%%%%%%%%%%%%%%%%%%%%%%%%%%%%%%%%%%%%%%%%%%%%%%%%%%%%%%%%%%%%%%%%%%%%%

\section{T-Duality and D-Branes}

Let us consider a closed string moving in the circle of radius $R$,
which is described by a periodic scalar field $X\equiv X+2\pi R$
on the worldsheet.
The space of states is decomposed into sectors
labeled by two conserved  charges
--- the momentum $l\in\Z$ associated with the translation symmetry
$X\to X+$constant, and the winding number $m\in\Z$ which counts how many times
the string winds around the circle.
The ground state in each sector has energy
${1\over 2}[(l/R)^2+(Rm)^2]-{1\over 12}$, which
is invariant under
\begin{equation}
R\longleftrightarrow {1\over R},
\nonumber
\end{equation}
and $l\leftrightarrow m$.
In fact, the sigma model on the cricle of radius $R$
is equivalent to the sigma model on the circle of radius $1/R$.
This is called {\it T-duality}.
The exchange of momentum and winding number can be described as the relation
between the corresponding currents
\begin{equation}
\partial_t X=\partial_{\sigma}\widetilde{X},
\quad
\partial_{\sigma} X=\partial_t\widetilde{X},
\label{Tcurr}
\end{equation}
where $(t,\sigma)$ are the time and space coordinates on the worldsheet
and $\widetilde{X}\equiv\widetilde{X}+2\pi/R$ is the coordinate
of the T-dual circle.

Let us now introduce an open string to this system.
We need to specify the boundary condition on the scalar field $X$
at the worldsheet boundary, say, at $\sigma=0$.
Neumann boundary condition $\partial_{\sigma}X|_{\sigma=0}=0$
corresponds to the freely moving end point,
while Dirichlet boundary condition
$\partial_tX|_{\sigma=0}=0$ fixes the end point.
They describe open strings ening on {\it D-branes}:
the former is for a D1-brane
wrapped on the circle while the latter
is for a D0-brane at a point of the circle.
By the relation (\ref{Tcurr}),
we see that T-duality exchanges the Neumann and Dirichlet boundary
conditions.
Thus, T-duality maps a D1-brane wrapped on the circle to
a D0-brane at a point of the T-dual circle.
The open string end point is charged under $U(1)$ gauge field on the
D-brane.
The holonomy $a=\int_{S^1} A$ parametrizes
 the gauge field configuration on the D1-brane wrapped on the circle.
Under T-duality, this parameter is mapped to the position of
the D0-brane in the T-dual circle.
Therefore, T-dual of $S^1$ can be identified
as the dual circle $H^1(S^1,U(1))$.
This story generalizes to the higher dimensional torus
$T^n$: T-duality inverts the radiii of the torus, mapping
a D$n$-brane wrapped on $T^n$ to a D0-brane at a point of the T-dual torus
$\widetilde{T}^n$, where the map provides the identification
$\widetilde{T}^n\cong H^1(T^n,U(1))$.

The same story applies to
{\it supersymmetric} theories on the worldsheet, which are obtained by
including fermionic fields as the superpartner of the scalar fields.
%There is a natural supersymmetric extention of the Neumann and
%Dirichlet boundary conditions.
When the target space is a Kahler manifold,
the system has {\it $(2,2)$ supersymmetry} which is an extended symmetry
with four supercharges $Q_+,\bQ_+,Q_-,\bQ_-$
($\pm$ shows the worldsheet chirality.
The supercharges are complex but are related by
hermitian conjugation $\bQ_{\pm}=Q_{\pm}^{\dag}$).
The complex coordinates of the target space
are annihilated by $\bQ_{\pm}$-variation, but are mapped
to the partner Dirac fermions under the $Q_{\pm}$-variation.
The simplest example is the cylinder
$\C^{\times}=\R\times S^1$ with the (flat) product
metric parametrized by the radius $R$ of the circle.
T-duality applied to the $S^1$ yields another cylinder
$\widetilde{\C}^{\times}=\R\times\widetilde{S}^1$
with radius $1/R$. This is actually an example of {\it mirror symmetry}:
$Q_-$ and $\bQ_-$ are exchanged under the equivalence.
%$Q_{\pm}$ variation of the complex coordinate of $\C^{\times}$
%yields the partner Dirac fermion, while
%it is the $Q_+,\bQ_-$ variation that 
%pairs the complex coordinate of $\widetilde{\C}^{\times}$
%to the Dirac fermion.
T-duality maps the various D-brane configurations:
D2-brane wrapped on $\C^{\times}$ is mapped to
D1-brane extedning in the $\R$-factor of $\widetilde{\C}^{\times}$;
D1-brane wrapped on $S^1$ at a point in the $\R$-direction
is mapped to D0-brane at a point of $\widetilde{\C}^{\times}$.

\begin{figure}[htb]
\centerline{\includegraphics{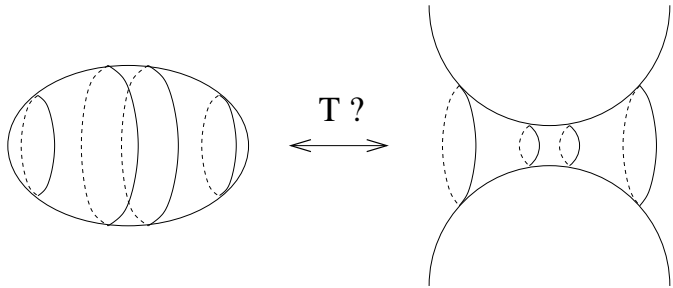}}
\caption{Is this what T-duality does?}
\label{t?}
\end{figure}

Let us next consider a more interesting target space ---
the two-sphere $S^2$.
It can be viewed as the circle fibration over a segment, and one may ask
what happens if T-duality is applied fiber-wise.
Since T-duality inverts the radius of the circle,
larger circle is mapped to smaller circle and smaller circle is mapped to\
larger circle, and one may na\"\i vely
expect that the dual geometry is as in Fig.~\ref{t?}.
Since the size of the dual circle blows-up
toward the two ends,
two holes effectively opens up
and the dual geometry has topology of cylinder.
This is consistent in one aspect:
the conserved momentum associated with
the $U(1)$-isometry of $S^2$ (fiber-rotaion) is mapped to the
winding number of the dual system, which is conserved due to
the cylindrical topology.
However, another aspect is not clear.
The winding number is not conserved in the
original sysetm because $\pi_1(S^2)=\{1\}$, and this should mean
in the dual theory that
the momentum is not conserved or the translation symmetry is broken.
But how can it be broken?
Is it because the metric is secretly not invariant under
rotation of the cylinder?

What really happens under T-duality is as follows.
It is true that the dual geometry has topology of cylinder.
However, the dual theory is not just a sigma model but a model called {
\it Landau-Ginzburg (LG) model}. It has a potential and
Yukawa coupling terms determined by a holomorphic function
of the target space, called the {\it superpotential}.
Let us parametrize the dual cylinder by a complex coordinate $Y$
which is periodic in the imaginary direction
$Y\equiv Y+2\pi i$.
Then, the superpotential of the dual LG model is given by
\begin{equation}
W=\e^{-Y}+\e^{-t+Y}.
\label{SG}
\end{equation}
Here $t=r-i\theta$
is a complex parameter 
that corresponds to the data
of the original $S^2$ sigma model:
$r$ is the area of the original $S^2$ and
$\theta$ determines the B-field
(it gives a phase factor $\e^{ik\theta}$ to the path-integral
measure for a worldsheet mapped to $S^2$ with degree $k$).
It is this superpotential that breaks the translation symmetry
${\rm Im}Y\to{\rm Im}Y+$ constant.
This T-duality is a mirror symmetry, as in the example of cylinder.

This was derived in \cite{HV} by exact analysis of quantum field theory
on the worldsheet.
The derivation applies to the case where the target space is a general
toric manifold $X$.
A toric manifold, as $S^2$ is, can be viewed as the torus fibration over some
base manifold, and T-duality sends the sigma model to a LG model. 
Suppose $X$ is realized as the symplectic quotient of $\C^N$ by
$U(1)^k$ action $(z_i)\mapsto (\e^{iQ_i^a\lambda_a}z_i)$
with the moment map equation
$\sum_{i}Q_i^a|z_i|^2=r^a$, and suppose the B-field is such that
the path-integral weight is $\e^{i\sum_ak_a\theta^a}$ for
 a map of multi-degree $(k_a)$.
Then, the dual geometry is an $(N-k)$-dimensional cylinder
defined by $\sum_iQ_i^aY_i=r^a-i\theta^a$ for $Y_i\equiv Y_i+2\pi i$,
and the superpotential is given by
$W=\sum_i\e^{-Y_i}$.
This mirror symmetry explains several observations made earlier
\cite{FI,Batyrev0,Giv0,EHY,EHX}.
The analysis of \cite{HV}
also includes the derivation of the mirror pairs
of Calabi-Yau hypersurfaces or complete intersections
in toric manifolds \cite{GP,Batyrev2}.
Furthermore, the method can be applied also to string backgrounds
with non-trivial dilaton \cite{HK} and H-field \cite{Tong}.

We do not repeat the analysis of \cite{HV} in what follows.
Instead, we will find some of the consequences of the mirror symmetry,
especially on D-branes.
We will see that D-brane analysis sheds new light on the duality itself.
Recall that, for the circle sigma model,
the dual circle was identified as the space of wrapped D1-branes,
$\widetilde{S}^1\cong H^1(S^1,U(1))$.
The same will happen here; the dual theory can be rediscovered by
looking at the D-branes wrapped on the circle fibers of the
toric manifold.

\section{Supersymmetric D-branes}\label{sec:DD}

\newcommand{\Bgamma}{\mbox{\large $\gamma$}}

Abstractly, D-branes
can be regarded as boundary conditions or boundary interactions
on the worldsheet of an open string.
We will focus on those preserving a half of
the $(2,2)$ worldsheet supersymmetry.
There are two kinds of such D-branes
\cite{OOY}:
A-branes preserving the combinations $Q_A=\bQ_++Q_-$ and
$Q_A^{\dag}=Q_++\bQ_-$;
B-branes preserving $Q_B=\bQ_++\bQ_-$ and $Q_B^{\dag}=Q_++Q_-$.
Since mirror symmetry exchanges $Q_-$ and $\bQ_-$,
A-branes and B-branes are exchanged under mirror symmetry.

Let us consider the sigma model on a K\"ahler manifold $M$.
$M$ can be considered as a complex manifold or as a symplectic manifold
(with respect to the K\"ahler form $\omega$).
A D-brane wrapped on a cycle $\Bgamma$ of $M$ and supporting
a unitary gauge field $A$ is an A-brane if $\Bgamma$
is a Lagrangian submanifold ($\omega|_{\gamma}=0$) and $A$ is flat ($F_A=0$),
while it is a B-brane if $\Bgamma$ is a complex submanifold of $M$ and
$A$ is holomorphic ($F_A^{(2,0)}=0$).
If we consider a LG model with superpotential $W$,
there is a further condition that
the $W$-image of $\Bgamma$ is a straightline parallel to the real axis
for A-branes and $W$ is a constant on $\Bgamma$ for B-branes
\cite{HIV,GJS}.
A-branes and B-branes
are objects of interest from the point of view of symplectic
geometry and complex analytic geometry, respectively.
They are exchanged under mirror symmetry.

Of prime interest are the lowest energy states
of open strings ending on D-branes.
In particular, the supersymmetric ground states
which correspond to massless open string modes.
The theory of an open string stretched between two A-branes (or two B-branes)
has one complex supercharge $Q=Q_A$ (or $Q=Q_B$).
In many cases, it obeys the supersymmetry algebra
\begin{equation}
\begin{array}{c}
\{Q,Q^{\dag}\}=2H,\\
Q^2=0,
\end{array}
\end{equation}
where $H$ is the Hamiltonian of the system.
Then, the system can be regarded as
the supersymmetric quantum mechanics
(with infinitely many degrees of freedom)
and the standard method \cite{SQM} applies.
In particular, there is a one-to-one correspondence
between the gound states and the $Q$-cohomology classes.
However, in some cases, the above algebra is modified
and it can happen that
\begin{equation}
Q^2\ne 0.
\end{equation}
In such a case, the cohomological characterization of the ground states
does not apply.
(In fact, there is no supersymmetric ground states.)
This does not happen for closed strings
and is a new phenomenon peculiar to open strings.

In what follows, we study D-branes in the sigma model on
$S^2$ or more general toric manifolds.
In Sec.~\ref{sec:AB}, we study A-branes in $S^2$
and the mirror B-branes in the LG model.
We will see that we can reproduce the mirror duality
through the study of D-branes.
In Sec.~\ref{sec:BA}, we study B-branes in $S^2$
and the mirror A-branes in LG.

\section{Intersecting Lagrangians and their Mirrors}\label{sec:AB}

Let $(M,\omega)$ be a Kahler manifold.
We study the open string stretched from
one A-brane $(\Bgamma_0,A_0)$ to another $(\Bgamma_1,A_1)$,
where $\Bgamma_i$ are Lagrangian submanifolds and $A_i$ are
flat $U(1)$ connections on them.
Classical supersymmetric
configurations are the ones
mapped identically to the intersection points of
$\Bgamma_0$ and $\Bgamma_1$.
However, quantum tunneling effects may lift the ground state degeneracy.
Only the index $\Tr(-1)^F=\#(\Bgamma_0\cap\Bgamma_1)$
is protected from corrections.

To determine the actual ground state spectrum,
one may apply Morse theory analysis of \cite{SQM}
to the space of open string configurations.
The sigma model action defines a Morse function
and its critical points are indeed the constant maps to
the intersection points of
$\Bgamma_0$ and $\Bgamma_1$.
Tunneling configurations are holomorphic maps
from the strip to $M$ such that
the left and the right boundaries are mapped to
$\Bgamma_0$ and $\Bgamma_1$ respectively,
and the far past and the far future 
are asymptotic to the constant maps
to $\Bgamma_0\cap\Bgamma_1$.
This usually leads to a cochain complex
that models the original $Q$-complex.
However,
the `coboundary' operator may fail to be nilpotent
\cite{FOOO}, $\partial^2\ne 0$,
which corresponds to $Q^2\ne 0$.

\begin{figure}[htb]
\centerline{\includegraphics{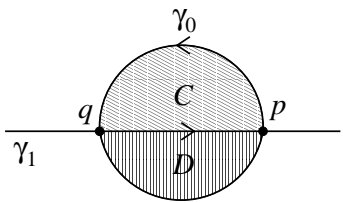}}
\caption{An example with $Q^2\ne 0$}
\label{cook}
\end{figure}

As an example, consider two Lagrangian submanifolds in the complex plain
$M=\C$ as depicted
in Fig.~\ref{cook}.
They intersect at two points $q$ and $p$, and the constant maps to them
are the candidate supersymmetric configurations.
The `cochain' complex has $\Z_2$ grading that distinguishes $p$ and $q$.
There is one tunneling configuration from $q$ to $p$ ---
the homolorphic map from the strip to the region $C$. Then,
the `coboundary' operator acts as
$\partial q=\e^{-A(C)}p$ where $A(C)$ is the area of the region $C$.
Likewise, we find
$\partial p=\e^{-A(D)}q$.
Then, we see that
\begin{equation}
\partial^2 q=\e^{-A(C)}\e^{-A(D)}q=\e^{-A(C\cup D)}q\ne 0.
\end{equation}
The standard proof of $\partial^2=0$ does not apply here:
there is a one parameter family of tunneling configurations from $q$ to $q$,
 that starts with the composition of $C$ and $D$ at $p$
and ends with the holomorphic disc $C\cup D$.
%(Let $r$ be a point on the segment $[q,p]$ in $\Bgamma_1$.
%Then, there is a holomorphic map form the strip to
%$C$ and $D$ glued along the segment $[r,p]$.
%This makes the family parametrized by the position of
%$r$.)
This is a general phenomenon called 
``bubbling off of holomorphic discs'', which is
peculiar to open string systems.
%and cannot happen for closed strings on dimensional ground.

If $\partial$ happens to be nilpotent, one can define the cohomology group,
which is known as the Floer cohomology group
$HF((\Bgamma_0,A_0),(\Bgamma_1,A_1))$.
This is the space of supersymmetric ground states of the open string system.

Let us next consider a LG model with
superpotential $W$ on a complex manifold $Y$. We study the open string
stretched from a B-brane $Z_0$ to another $Z_1$, where $Z_i$ are complex
submanifolds of $Y$ on which $W$ are constants.
It is straightforward to show, using the canonical commutation relation,
that
\begin{equation}
Q^2=W|_{Z_1}-W|_{Z_0},
\label{Q2}
\end{equation}
and there is no quantum correction to it.
Thus, we see that $Q^2\ne 0$ if the $W$-values of $Z_0$ and $Z_1$
do not agree.
If they do agree, the space of supersymmetric ground states
is the $Q$-cohomology group.
There is a finite dimensional model of the $Q$-complex
\cite{H,HKKPTVVZ};
It consists of anti-holomorphic forms on
$Z_0\cap Z_1$ with values in the exterior powers of
$N_{Z_0}\cap N_{Z_1}$,
on which the coboundary operator acts as
$\bartial+\partial W\cdot$
(here $N_{Z_i}$ is the normal bundle of $Z_i$ in $Y$, and
$\partial W\cdot$ is the contraction with the holomorphic
1-form $\partial W$).
If $Z_0$ and $Z_1$ are points, the complex is non-trivial
only if they are the same point, and the cohomology
is the exterior power of the tangent space if the point is a critical point of
$W$ but vanishes if the point is not a critical point.

Eqn.~(\ref{Q2}), if non-zero, is the mirror counterpart of
$Q^2\ne 0$ for the intersecting Lagrangian systems.
Note that we find (\ref{Q2}) by purely
classical analysis, in contrast to the case of A-branes
where computation of $Q^2$ requires the analysis of quantum tunneling
effect.

To make it explicit, let us come back to the mirror symmetry between
the $S^2$ sigma model and the LG model with superpotential
$W=\e^{-Y}+\e^{-t+Y}$.
We consider A-branes wrapped on the $S^1$ fibers in the Fig.~\ref{t?}(Left).
They are mapped under T-duality to B-branes at points on
the $Y$-cylinder ---
their ${\rm Re}(Y)$ and ${\rm Im}(Y)$
coordinates are determined respectively by the location of the $S^1$
and the holonomy of the $U(1)$ gauge field.
One can actually find a detailed map via a field theory analysis \cite{H}.
Let $Y=c-ia$ be the location of the mirror D0-brane.
Then, the area of the disc bounded by the original D1-brane
is $c$ 
and the holonomy of the $U(1)$ connection is
$\e^{i(a-c\theta/r)}$.
Let us now find the condition for $Q^2=0$ using the LG description.
For two D0-branes at $Y=c_0-ia_0$ and
$Y=c_1-ia_1$,
we find
\begin{equation}
Q^2=\e^{-c_1+ia_1}+\e^{-t+c_1-ia_1}
-\e^{-c_0+ia_0}-\e^{-t+c_0-ia_0}.
\label{LGcond}
\end{equation}
There are two solutions to $Q^2=0$:
$(c_1,a_1)=(c_0,a_0)$ and $(r-c_0,\theta-a_0)$.
In the $S^2$ side, 
they correspond to two identical D1-branes (the same location
and the same holonomy),
and two D1-branes of opposite holonomies
such that the inside area of one is equal to the outside
area of the other.
The $Q$-cohomology is non-trivial only if the two points are
the same critical point of the superpotential $W$, which is
$\e^{-Y}=\e^{-t/2}$ or $-\e^{-t/2}$ in the present case.
In such a case, the cohomology is
the exterior power of the tangent space,
$\wedge\C=\wedge^0 \C\oplus\wedge^1\C$,
which has one bosonic and one fermionic
basis vectors.
In the original $S^2$ sigma model,
$\e^{-c+ia}=\pm\e^{-t/2}$
means that the inside and outside area of $S^1$ are the same
and the holonomy is $\pm 1$.
Thus, we find that mirror symmetry predicts
\begin{equation}
HF((\Bgamma_0,A_0),(\Bgamma_1,A_1))
=\left\{
\begin{array}{ll}
\wedge\C
&\begin{array}{l}
\mbox{if $\Bgamma_0=\Bgamma_1$ divides $S^2$ into ${1\over 2}$}\\
\mbox{~~and $A_0=A_1$ has holonomy $\pm 1$}
\end{array}
\\[0.3cm]
0
&\mbox{otherwise}
\end{array}\right.
\label{LGpre}
\end{equation}

\begin{figure}[htb]
\centerline{\includegraphics{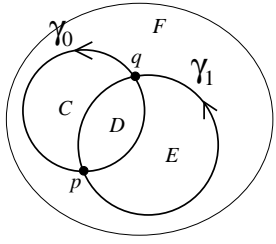}}
\caption{Intersecting Lagrangians in $S^2$}
\label{bS2}
\end{figure}

One can also directly compute the Floer homology group.
Let us put the D1-branes in a position as
in Fig.~\ref{bS2}.
For simplicity, we set off the $\theta$-angle as well as
the $U(1)$ connections on $\Bgamma_i$.
The two circles intersect at two points $q$ and $p$
which represent `cochains' of different degrees
(the `complex' is $\Z_2$ graded).
The `coboundary' operator acts as
$\partial q=\e^{-A(D)}p-\e^{-A(F)}p$
and $\partial p=\e^{-A(C)}q-\e^{-A(E)}q$.
Thus the square is
$\partial^2q=(\e^{-A(C\cup D)}-\e^{-A(D\cup E)}
-\e^{-A(C\cup F)}+\e^{-A(E\cup F)})q$.
If we denote the region inside
$\Bgamma_i$ by $D_i$, the area of the region outside
$\Bgamma_i$ is $r-A(D_i)$.
We therefore find
 \begin{equation}
\partial^2 q=\left[
\e^{-A(D_0)}-\e^{-A(D_1)}
-\e^{-r+A(D_1)}+\e^{-r+A(D_0)}\right]q
\label{FHcond}
\end{equation}
This vanishes if and only if $A(D_1)=A(D_0)$ or $A(D_1)=r-A(D_0)$,
namely, when the inside area of $\Bgamma_1$ is equal to the inside
or outside area of $\Bgamma_0$.
One can also show that the $U(1)$ holonomy of the two should be
the same or opposite, respectively.
This matches precisely with the LG result.
In fact, (\ref{FHcond}) with holonomy included
is identical to (\ref{LGcond})
under the map of variables mentioned before.
Let us next compute the cohomology.
Consider $A(D_0)=A(D_1)$ first.
In such a case, $A(C)=A(E)$ and 
$A(F)-A(D)=A(F\cup C)-A(D\cup C)=r-A(D_1)-A(D_0)=r-2A(D_0)$.
Thus the coboundary operator acts as
\begin{equation}
\begin{array}{l}
\partial q=\e^{-A(D)}(1-\e^{-r+2A(D_0)})p\\
\partial p=0.
\end{array}
\nonumber
\end{equation}
The cohomology vanishes if $A(D_0)\ne r/2$, while
it is non-vanishing if $A(D_0)= r/2$
--- each of $q$ and $p$ generates the cohomology at its degree.
Note that $A(D_0)=r/2$ is when $\Bgamma_0$ divides $S^2$ into half.
One can also show that the cohomology is non-vanishing if the
holonomy is $\pm 1$.
We also find the same conclusion 
if we start with $A(D_1)=r-A(D_0)$.
To summarize,
we see that the result (\ref{LGpre}) of mirror symmetry
is indeed correct.

We have seen a practical aspect of mirror symmetry
in the study of D-branes.
Computation of ground state spectrum of open string
involves highly non-trivial analysis of quantum tunneling effect in
the sigma model,
while it is done by a simple classical manipulation in the mirror LG
model.
This story generaizes straightforwardly to
more general toric manifolds.

There is another important aspect in the above analysis.
We recall that, in the case of $S^1$ sigma model,
it was enough to analyze the wrapped D1-brane to find
the T-dual space; since D1-brane is T-dual to D0-brane,
its moduli space is equivalent to
the space of D0-branes of the T-dual theory, namely the dual space itself.
In fact, the same applies here as well.
By analysing the D1-branes wrapped on the $S^1$-fibers,
we find the cylinder as the dual space, but we also find
the superpotential (up to constant addition)
through the computation of $Q^2$,
see (\ref{FHcond}).
In other words, we can reproduce the mirror symmetry
between toric sigma models and LG models
by analyzing the D-branes.

This point of view shares its spirit with
Strominger-Yau-Zaslow \cite{SYZ} who
proposed, using D-branes,
that mirror symmetry of Calabi-Yau manifolds
is nothing but dualization of special Lagrangian fibrations.
The latter has led, for example,
to topological construction of mirror manifolds \cite{Ruan,Gross,Morrison}.
In attempts to make it more precise,
the treatment of singular fibers constitues the essential
part where quantum corrections are expected to play an important role.
(See e.g. \cite{Fukaya} for a recent progress.)
The example considered above includes singular
fibers and we have shown how the quantum effect is taken into account.
Although we have not dealt with {\it special} Lagrangian fibrations,
we note that the above analysis applied to
toric Calabi-Yau yields mirror manifolds
cosistent with SYZ program.
For example, if we start with the total space of
${\mathcal O}(-1)\oplus{\mathcal O}(-1)$ over $\CP^1$,
we obtain the LG model on $(\C^{\times})^3$ with superpotential
$W=\e^{-Y_0}(\e^{-Y_1}+\e^{-Y_2}+\e^{-t-Y_1-Y_2}+1)$ as the mirror,
which in turn is related \cite{HIV}
to the sigma model on the Calabi-Yau hypersurface
$\e^{-Y_1}+\e^{-Y_2}+\e^{-t-Y_1-Y_2}+1=uv$ in
$\C^{\times}\times\C^{\times}\times\C\times\C=\{(\e^{-Y_1},\e^{-Y_2},u,v)\}$.
This last mirror turns out to be consistent 
with SYZ topologically \cite{Gross2}.

\section{Holomorphic Bundles and their Mirrors}
\label{sec:BA}

Let us consider an open string stretched between B-branes
wrapped on $M$ and supporting holomorphic vector bundles $E_0$ and $E_1$.
The zero mode sector of the open string Hilbert space
is identified as the space
$\Omega^{0,\bullet}(M,E_0^{*}\otimes E_1)$ where the supercharge
$Q$ acts as the Dolbeault operator.
Thus, the space of supersymmetric ground states
in the zero mode approximation is the Dolbeault cohomology
or ${\rm Ext}^{\bullet}(E_0,E_1)$.
In particular, the index is
\begin{equation}
\Tr(-1)^F=\chi(E_0,E_1).
\nonumber
\end{equation}
In the full theory, some pairs states
of neighboring R-charges could be lifted to non-supersymmetric states.
The latter does not happen if ${\rm Ext}^{p}(E_0,E_1)$ is non-zero
only for even $p$ (or odd $p$).
An example of such a pair $E_0,E_1$ is
from an {\it exceptional collection} \cite{rudakov},
which is an ordered set of bundles $\{E_i\}$
where ${\rm Ext}^p(E_i,E_i)=\delta_{p,0}\C$
while for $i<j$ ${\rm Ext}^{\bullet}(E_j,E_i)=0$ but
${\rm Ext}^{p}(E_i,E_j)$ can be non-zero
only for one value of $p$.
For $\CP^n$, the set of line bundles
$\{{\mathcal O}(i)\}_{i=j}^{j+n}$
is an exceptional collection ($\forall j\in\Z$).

Next we consider the LG model with superpotential $W$
which has only non-degenerate critical points $\{p_i\}$.
Gradient flows of ${\rm Re}\, W$ originating from $p_i$ sweep
a Lagrangian submanifold $\Bgamma_i$ whose $W$-image is a
straight horizontal line emanating from the critical value
$w_i=W(p_i)$. Thus, the D-brane wrapped on $\Bgamma_i$ is an A-brane.
Let us consider an open string stretched from
$\Bgamma_i$ to $\Bgamma_j$.
Classical supersymmetric configurations
are gradient flows of
$-{\rm Im}\,W$ from a point in $\Bgamma_i$ to
a point in $\Bgamma_j$.
The index is the number of such gradient flows
counted with an appropriate sign.
It is
\begin{equation}
\Tr(-1)^F
=\#(\Bgamma_i^-\cap\Bgamma_j^+)
\nonumber
\end{equation}
where $\Bgamma_k^{\pm}$
is the deformation
of $\Bgamma_k$ so that the $W$-image
is rotated at $w_k$ with a small angle $\pm \epsilon$.
Quantum mechanically, the paths of opposite signs are lifted
by instanton effects and
$|\#(\Bgamma_i^-\cap \Bgamma_j^+)|$
 is in fact the number of
supersymmetric ground states.
One can also quantize the system
using the Morse function determined by the LG action.
This leads to the LG version of the Floer homology group
${\rm HF}_{W}^{\bullet}(\Bgamma_i,\Bgamma_j)$.
(This is studied also by Y.-G.~Oh \cite{Oh}.)
If ${\rm Im}w_i>{\rm Im}w_j$ and there is no critical value between
the $W$ images of $\Bgamma_i$ and $\Bgamma_j$,
then $\#(\Bgamma_i^-\cap \Bgamma_j^+)$
is equal to the number $S_{ij}$ of BPS solitons
connecting $p_i$ to $p_j$.

B-branes supporting holomorphic bundles on
toric manifolds are mapped under mirror symmetry to
A-branes in the LG models.
Let us examine the detail in the $S^2$ sigma model and its mirror
LG model $W=\e^{-Y}+\e^{-t+Y}$.
The superpotential has two critical points $p_{\pm}$ with
critical values $w_{\pm}=\pm 2\e^{-r/2+i\theta/2}$.
We first set $\theta=0$ so that $w_{\pm}$ are on
the real line.
The simplest brane on $S^2$ is the $U(1)$-bundle with trivial connection,
or ${\mathcal O}$ over $\CP^1$.
Since we are T-dualizing along the $S^1$-fibers
along which the holonomy is trivial,
the dual has to be localized at a constant point in the dual fibers.
It is a D1-brane at the horizontal line ${\rm Im}Y=0$ whose
$W$-image is a straigtline emanating from $w_+$ --- the A-brane
$\Bgamma_{+}$.
Another simple B-brane is the D0-brane at a point.
Its dual is wrapped on $S^1$ but remains localized in the horizontal
direction. The one at ${\rm Re}Y=r/2$ has straight $W$-image
and therefore is an A-brane, which we call $\Bgamma_0$.
\begin{figure}[htb]
\centerline{\includegraphics{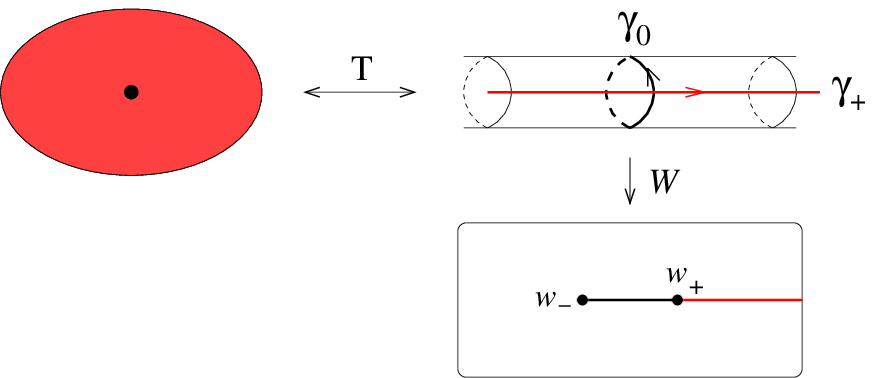}}
\caption{}
\label{BA}
\end{figure}
Introducing $n$ D0-branes to ${\mathcal O}$ means turning on $n$ units of
magnetic flux and produces the brane ${\mathcal O}(n)$.
Its dual is the combination of $\Bgamma_{+}$ and
$n$ $\Bgamma_0$. For negative $n$, $n\Bgamma_0$ is understood to have
the reversed orientation.

\begin{figure}[htb]
\centerline{\includegraphics{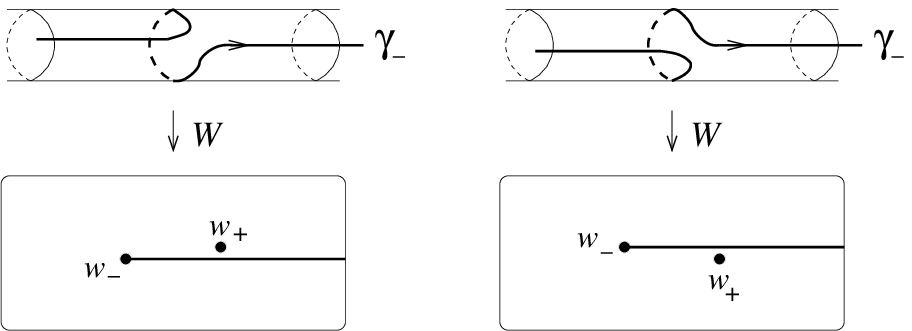}}
\caption{}
\label{CC}
\end{figure}

Let us now turn on a small positive $\theta$. Then, $w_+$ is slightly above
$w_-$ in the imaginary direction, and $\Bgamma_0$
is no longer an A-brane. Instead
we find a non-compact A-brane $\Bgamma_-$.
See Fig.~\ref{CC}(Left).
$\Bgamma_-$ belongs to the homology class $\Bgamma_++\Bgamma_0$
and therefore is the mirror of ${\mathcal O}(1)$.
We note that indeed $\chi({\mathcal O},{\mathcal O}(1))
=\#(\Bgamma_+^-\cap\Bgamma_-^+)=2$.
If we turn on a small negative $\theta$,
then $w_+$ is slightly below $w_-$, and we find the A-brane $\Bgamma_-$
in the homology class $\Bgamma_+-\Bgamma_0$ (Fig.~\ref{CC}(Right)).
Since the orientation
of $\Bgamma_0$ is reversed, $\Bgamma_-$ is the mirror of
${\mathcal O}(-1)$ for this value of $\theta$.

Note that $\Bgamma_-$ for $\theta=+\epsilon$ and
$\Bgamma_-$ for $\theta=-\epsilon$
are related by Picard-Lefshetz formula
$\Bgamma_-|_{\theta=+\epsilon}
=(-\Bgamma_-+2\Bgamma_+)|_{\theta=-\epsilon}$
where the coefficient $2$ is $\#(\Bgamma_+^-\cap\Bgamma_-^+)$
at $\theta=+\epsilon$.
On the other hand,
the mirror bundles ${\mathcal O}(1)$ and ${\mathcal O}(-1)$
appear in the exact sequence
\begin{equation}
0\longrightarrow
{\mathcal O}(-1)\longrightarrow
{\rm Ext}^0({\mathcal O},{\mathcal O}(1))\otimes {\mathcal O}
\longrightarrow
{\mathcal O}(1)
\longrightarrow 0.
\nonumber
\end{equation}
In such a case, ${\mathcal O}(-1)$ is said to be the {\it left mutation} of
${\mathcal O}(1)$ with respect to
${\mathcal O}$ \cite{rudakov}.
It was observed in \cite{zaslow,Kon2} (see also \cite{Dubro})
that mutation of exceptional bundles in certain Fano manifolds
is related to Picard-Lefshetz monodromy that appears in the
theory of BPS solitons in the sigma models.
We can now understand it as a consequence of mirror symmetry
in the case where the taget space is a toric manifold.
Related works has been done by P.~Seidel \cite{Seidel}.

\section{Concluding Remarks}

We have presented some applications of mirror symmetry
between toric manifolds and LG models \cite{HV}, especially
to the study of D-branes.
In particular,
we have seen that D-branes wrapped on torus fibers
can tell us about the mirror symmetry itself.
Below, we comment on some materials that are not covered here.

Structure of integerable system
in topological string theory, along with the
matrix model representation,
is an important aspect of quantum geometry.
It was first discovered in topological gravity
\cite{Witten,Kon1},
the case where the target space is a point.
There are several observations suggesting
that it may extend to more general target spaces.
Here the mirror LG superpotential is expected to play an
important role.
For example, for $\CP^1$ model,
$W=\e^{-Y}+\e^{-t+Y}$ is the Lax operator of
the Toda lattice hierarchy \cite{EY,EHY} under the replacement of
$Y$ by a differential operator
(see also \cite{Getzler,PP}).
For a projective space of higher dimension,
the mirror superpotential also plays the role of
Lax operator at least in genus zero \cite{EHX}.
Also, the Virasoro constraint \cite{EHX2}
suggests existence of matrix model reprtesentations
which in turn are related to integrable systems.
Some beautiful story is waiting there to be discovered.
Possible role of branes is interesting to explore.

Mirror symmetry has an application to enumerative geometry
including holomorphic curves with boundaries.
In the case of special Lagrangian submanifolds in Calabi-Yau three-folds,
the number of holomorphic curves ending on the submanifolds
enters into certain terms in the low energy effective action of
the superstring theory.
The number of holomorphic discs for a class of
special Lagrangians in toric Calabi-Yaus are counted in \cite{mina}
by computing the space-time superpotential terms in the mirror side.
There are several related works including the mathematical
tests \cite{LK,GZ} of \cite{mina}.

The derivation of \cite{HV} itself
applies only to toric manifolds and submanifolds therein
that are realized as vacuum manifolds of {\it abelain} gauge systems.
There are some observations that suggest the form of LG-type mirror for
Grassmann and flag manifolds \cite{EHX,Giv} which are realized as
the vacuum manifolds of {\it non-abelain} gauge systems.
It would be a challenging problem to find or derive
the mirror of such manifolds and others.
The method presented here using D-branes is possibly of some use.
Another possible way is to consider compactification
of three-dimensional mirror symmetry \cite{IS}.
This works for the abelain gauge systems \cite{AHKT}
and many examples of non-abelian mirror pairs are known
in three-dimensional gauge theories (e.g. \cite{DHOO}).

\section*{Acknowledgement}

I would like to dedicate this manuscript
to the memory of Dr. Sung-Kil Yang whom I respect with no limit.
I wish to thank
T.~Eguchi, A.~Iqbal,
C.~Vafa, S.-K.~Yang himself and C.-S.~Xiong for collaborations in
\cite{EHY,EHX,HV,HIV}.
I would also like to thank K.~Fukaya, Y.-G.~Oh,
H.~Ohta and K.~Ono
for useful discussions and explanation of their works.
This work was supported in part by NSF PHY 0070928.

\bibliographystyle{amsalpha}

\end{document}